\documentclass[12pt,epsbox]{article}
\usepackage{authblk}
\usepackage{hyperref}
\usepackage{color}
\usepackage[dvips]{graphicx}
\usepackage{amsmath}

\title{Notes on the Exact RG equation and the Wheeler-DeWitt equation}

\newcommand{\be}{\begin{equation}}
\newcommand{\ee}{\end{equation}}
\newcommand{\bs}{\begin{split}}
\newcommand{\es}{\end{split}}
\newcommand{\bea}{\begin{eqnarray}}
\newcommand{\eea}{\end{eqnarray}}
\newcommand{\ba}{\begin{eqnarray}}
\newcommand{\ea}{\end{eqnarray}}

\newcommand{\beq}{\begin{equation}}
\newcommand{\eeq}{\end{equation}}
\newcommand{\beqa}{\begin{eqnarray}}
\newcommand{\eeqa}{\end{eqnarray}}
\newcommand{\beqar}{\begin{eqnarray*}}
\newcommand{\eeqar}{\end{eqnarray*}}
\numberwithin{equation}{section}

\begin{document}

  \begin{flushleft}
    {\bf OUJ-FTC-11}
  \end{flushleft}
  
  \begin{center}
    \vspace{37pt}
    {\LARGE Notes on the Exact RG equation and the Wheeler-DeWitt equation}\\[1em]
    \vspace{17pt}
    {\large Hideto Kamei\footnote{hidetokamei@gmail.com}}\\[0.5em]
    \vspace{17pt}
    {Field theory group, \\
  The Open University of Japan,\\
  Chiba 261-8586, Japan}\\[1em]
    \vspace{17pt}
    {\large February 25, 2024}
  \end{center}
  \vspace{17pt}

In this note, in the context of the AdS/CFT correspondence, the holographic derivation of the Wilsonian effective action is proposed. Then, the exact RG equation in the boundary theory is derived from the Wheeler-DeWitt equation of the bulk, following the suggestion of \cite{dVV},\cite{VV}, and\cite{Polchinski}. The relationship between the exact RG and Stochastic Quantization\cite{Parisi},\cite{Damgaard} is briefly discussed.

\section{Introduction}
Here in this note, we would like to promote Holographic RG to exact RG, which corresponds to quantum gravity in the bulk.
By Verlinde et al \cite{dVV}, it was shown that the RG flow equation of the CFT can be derived from classical Equation of motion of the bulk\footnote{For a review, please refer to \cite{Sakai}.}. We generalize the RG flow equation to exact RG, which is derived from Wheeler-Dewitt equation in the bulk. We identify the expression of the boundary action and counterterm is properly subtracted in the derivation.

Now we outline the structure of this note. In section 2, the setup of the derivation is given. In section 3, the role of the boundary condition in the setup is explained. In section 4, procedure for splitting the bulk action and identifying the boundary action is given. In section 5, derivation of the exact RG equation from the bulk is given. In section 6, interpretation of the equation as the exact RG is given. In section 7, an example to show the validity of the methodology is given. In section 8, discussion on the result and possible future research is given. In section 9, update on research since this note was first written is given.



\section{Deriving the exact RG from the bulk theory }


Here we wish to derive the exact renormalization group equation (In the form of \cite{Morris},\cite{Morris2},\cite{Morris3},\cite{Oliver}) of $\mathcal{N}=4$ SYM from the Wheeler-DeWitt equation in the AdS. Most of the discussion follows from the beautiful paper on the holographic renormalization, \cite{dVV}. In the paper, the Callan-Symanzik type of the RG flow equation was derived from the condition that the Hamiltonian is zero, $H=0$, and the Wheeler-DeWitt equation is the quantum extension of $H=0$.

We assume that the bulk action contains gravity and a scalar field (the extension to including many scalar fields and including other fields is straightforward).
We take the parametrization of the 5d metric as,
\begin{equation}
ds^2 = g_{\mu\nu} dx^{\mu}dx^{\nu} = N^2 dr^2 +h_{ij}(dx^i +N^i dr)(dx^j +N^j dr).
\end{equation}

We will work in the Euclidean signature everywhere in this note.
Here we fix the gauge of the metric as $N=1, N_i=0$. 
 We decompose the Ricci scalar using extrinsic curvature, 
\begin{equation}
K_{ij} \equiv \frac{1}{2N} \left( \partial_r h_{ij} -D_i N_j -D_j N_i \right)=\frac{1}{2}\left( \partial_r h_{ij} \right).
\end{equation}

From now on we treat the metric as fields and use the condensed notation, $h_{ij} \equiv h^I ,K_{ij} \equiv K^I$.
Further we will use so-called DeWitt's supermetric $G_{IJ} \equiv G^{ijkl} \equiv \sqrt{h}\left( \frac{1}{2} (h^{ik}h^{jl}+h^{il}h^{jk})-h^{ij}h^{kl} \right)$ and its inverse $G^{IJ}$ for the condensed notation of the action.
We can define the conjugate momentum of the metric and the scalar field as $\pi_{I}=\delta \mathcal{L} / \delta \partial_r h^I$,
and $p=\delta \mathcal{L} / \delta \partial_r \phi$, and construct the Hamiltonian $H=\pi_I \partial_r h^I + p \partial_r \phi-\mathcal{L}$.
The equation we wish to solve is the Wheeler-DeWitt equation, which is $H \Psi =0$. Analogous to the Shr\"odinger equation, we simply have to replace momentum $\pi_I$, $p$ with the derivative $\delta / \delta h^I$, $\delta / \delta \phi$. We wish to proceed following \cite{VV} to derive the exact RG equation of the CFT holgraphically.
In the classical case, we will get the expression of the radial velocity of the fields (with respect to the 4d action) from Hamilton-Jacobi equation,

\begin{equation}
\begin{split}
\frac{\delta S_{UV}}{\delta \phi} &= \partial_r \phi\\
\frac{\delta S_{UV}}{\delta h^I}  &=  \frac{1}{16\pi G_N} G_{IJ} \partial_r h^{J} \label{velocity},
\end{split}
\end{equation}
and in the quantum case the same equation holds for field redefinition, but we will need a correction term $\mathcal{F}(\phi)$ as in (\ref{fieldtransf}), which will be consistent with Wheeler-DeWitt equation (\ref{HERG1}) we will derive.

\section{The role of the boundary condition and the fluctuation}

Here in this section, we would like to explain the role of the boundary condition in terms of the fluctuation in the boundary CFT, in order to motivate the holographic definition of the Wilsonian effective action in the following section.

First of all, we would like to claim that the boundary value of the bulk fields themselves could be considered as the dynamical fields in the boundary CFT in some cases. The main example we consider in this note is a scalar field in the AdS within the mass range $-d^2/4 < m^2 < -(d^2/4)+1$.
In this mass range, it is known, in addition to the standard Dirichlet boundary condition, we can also impose free boundary condition\cite{Witten2},\cite{Muck},\cite{Witten},\cite{WittenNeumann},\cite{MukundHRG}, and it is called "alternative quantization". Those two different boundary conditions correspond to different CFTs on the boundary, $CFT_D$ and $CFT_F$ respectively. In the case of free boundary condition, we also need to integrate over the value of the boundary condition of the bulk field, and therefore, the partition function of the $CFT_F$ can be obtained by promoting the source in the $CFT_D$ to a dynamical field and integrating over it as
\begin{equation}
\mathcal{Z}_F =\int \mathcal{D}J \mathcal{Z}_D = \int \mathcal{D}J \exp(-\Gamma[J])\label{FD}.
\end{equation}

This expression can be interpreted as the path integral formally (or statistical partition function if one wishes), where $J$ is now a dynamical field, and weight is $\exp(-\Gamma[J])$, $\Gamma$ being the generating functional for $CFT_D$. Therefore, in the case of the free boundary condition, the boundary value of the bulk field can be considered as the field in the boundary CFT itself, and we can consider an exact RG with respect to that.\footnote{Necessity of the free boundary condition was first realized by the necessity of having operator with small conformal dimension, which can't be realized by the naive Dirichlet boundary condition. Later it was realized those two CFTs are connected by an RG flow, by adding relevant perturbation $f \mathcal{O}^2$, where $\mathcal{O}$ is an operator to couple the source. (\ref{FD}) was obtained using that relationship.} In addition, also for other fields, it is known that in the free boundary condition (or equivalently Neumann boundary condition in the classical limit), the bulk fields on the boundary are interpreted as the dynamical fields in the boundary CFT.\cite{MukundOfer},\cite{Ross}.

Another example for the free boundary condition is, trailing string in the AdS\cite{StochasticdeBoer},\cite{StochasticSon}. We consider a string stretching from the horizon of the black brane to the boundary at the finite radial distance. The endpoint on the boundary is identified as the position of a particle in the CFT. Since the boundary is at the finite radial distance, the fluctuation of the string in the bulk can be transferred to the fluctuation of the endpoint, which is the fluctuation of the particle in the CFT. Therefore, we need to sum over all the possible value of the boundary position of the string, and it is formally the same as the path integral for the position of the particle in the CFT. Therefore, in the case of the trailing string, we also take the free boundary condition, and the boundary condition is promoted to a dynamical field in the boundary. we expect, if we take the boundary at the finite radial distance\cite{VV}, boundary value of the closed string modes in the bulk might be also identified as the fields of the closed string modes of the boundary theory.

Therefore, we claimed that, in some cases in the AdS/CFT, free boundary condition naturally arise and the integration over the boundary condition could be formally identified as the path integral in the boundary CFT. The above discussion will not be precise enough, but hopefully enough to motivate the following holographic definition of the Wilsonian effective action.

\section{The definition of the boundary Wilsonian effective action from the bulk}

When we define the Wilsonian effective action of the CFT side, we observe the CFT side with respect to a particular energy scale. In the context of AdS/CFT, it would be natural to concentrate on a hypersurface in the AdS, which will manifest the physics of the CFT side with respect to a particular energy scale we observe the physics \cite{dVV}. There the radial coordinate in the AdS is related to the energy scale of the observation in the CFT side. 
Let's take the hypersurface as $r=const$. Then, following \cite{Polchinski}, the total partition function of the bulk, which is also equal to the total partition of the CFT, can be written as the path integral over the fields on this boundary, and fields in the region of larger $r$ than this hypersurface (I denote it as UV, since it corresponds to the UV degree of freedom in the CFT side), and fields in the smaller $r$ region (I denote it as IR). Then, we can separate the expression of the bulk gravity action into UV part and IR part, since there is no cross term in the action for the fields with different $r$. Note that they are still the function of the Dirichlet boundary condition on the hypersurface.

 \begin{equation}
 \begin{split}
 \mathcal{Z} &= \int \mathcal{D}\phi_{UV} \mathcal{D} \phi_{IR} \mathcal{D} \phi_r \exp(-S^{bulk}_{UV}) \exp(-S^{bulk}_{IR})\\
             &= \int \mathcal{D}\phi_r \Psi_{UV}(r,\phi_r) \Psi_{IR}(r,\phi_r)\\
             &= \int \mathcal{D}\phi_r \exp(-S(r,\phi_r))\label{WilsonianPartition}
 \end{split}
 \end{equation}
 We can first perform the path integral for the fields of the UV and the IR part, and it will give the result of the second line. We expressed the UV part of the partition function, or wave function as $\Psi_{UV}(r,\phi_r)= \exp(-S^{bulk}_{UV}(r,\phi))$, and IR part of the wave function as $\Psi_{IR}(r,\phi_r)= \exp(-S^{bulk}_{IR}(r,\phi))$. Further, we can define the total low energy effective action as\\
 $\exp(-S(r,\phi_r)) \equiv \Psi_{UV}(r,\phi_r)\Psi_{IR}(r,\phi_r)$, which is the function of the boundary condition at $r$. The last line of (\ref{WilsonianPartition}) looks like the expression of the CFT partition function, with respect to the effective field $\phi_r$ and Wilsonian effective action $S(r,\phi_r)$, and it is tempting to define the Wilsonian effective action at some energy scale, which is determined by $r$, as simply this $S(r,\phi_r)$, holographically. Indeed, defining this way, we can show this "Wilsonian effective action" will follow the Exact RG equation with respect to the change of the energy scale, which is the boundary position $r$ in the bulk. This is given by solving the Wheeler-DeWitt equation in the bulk, which is the quantum extension of the condition that the Hamiltonian is zero for the gravity. Also note, that the above definition of the Wilsonian effective action, reduces to the definition of \cite{MukundHRG} in the bulk classical limit, as it should.

\section{Holographic Exact RG at finite radial distance}


Let's then derive the exact RG equation from the WDW equation in the bulk following the spirit of \cite{VV}. To make a contact with Polchinski type\cite{PolchinskiRG} of the Exact RG, let's first consider the regime where we can treat the metric as the classical background, and concentrate on the quantized scalar field in the AdS. This is done by recovering the Newton's constant $G_N$ to the Wheeler-DeWitt equation, and expanding it with $G_N$. This is done following \cite{GiladSchwinger}.

The Wheeler-DeWitt equation is,
\begin{equation}
\left[ 16 \pi G_N G^{IJ} \frac{\delta^2}{\delta h^I \delta h^J} - \frac{\sqrt{h}}{16\pi G_N} \left( (R-V) + (16\pi G_N) H_{matter} \right)  \right] \Psi=0.
\end{equation}

Both UV and IR wavefunction, $\Psi_{UV}$ and $\Psi_{IR}$ are expected to satisfy the above WDW equation separately.
Here we assume that $V$ is a constant (which doesn't depend on $\phi$), and $H_{matter}= (p^2/2) + W(\phi) -\Phi(\phi)R + (1/2)(\partial_i \phi)^2$, where $W(\phi)$ is the $\phi$ dependent part of the 5d potential, and its order $\phi^2$ has the coefficient $m^2/2$ which satisfies the bound that we discussed. We assume that $H_{matter}$ is the order of $\mathcal{O}(G_N^0)$. We further assume that we can expand
$S=-\log \Psi$ as $S=(S^0 / 16 \pi G_N) + S^1 + \mathcal{O}(G_N^1)$, and suppose $S_0$ depends only on the metric, and not on the scalar field. By expanding the Wheeler-DeWitt equation for $\mathcal{O}(G_N^{-1})$, while still seperating $S_0$ into the UV part and the IR part with respect to the cutoff, we get

\begin{equation}
\frac{1}{16 \pi G_N} \left(\frac{\delta S^0_{UV}}{\delta h} \right)^2 - \left( \frac{\delta^2 S^0_{UV}}{\delta h^2} \right) = -\frac{1}{16 \pi G_N} (\sqrt{h}) (V-R) =
\frac{1}{16 \pi G_N} \left(\frac{\delta S^0_{IR}}{\delta h} \right)^2 - \left( \frac{\delta^2 S^0_{IR}}{\delta h^2} \right).
\end{equation}

Note that the indices of the metric $h^I$ are implicitly contracted in the above expression. Strictly speaking, $\delta^2 S / \delta h^2$ term is order $\mathcal{O}(G_N^0)$, but we added it for simplicity of the next order \cite{GiladSchwinger}. This form of the 5d potential implies that the bulk gravity could be constructed out of the Stochastic Quantization\footnote{Please refer to \cite{Nishioka} for an analogous discussion.}, $ -\partial_r h^I=-G^{IJ}(\delta S_{UV} / \delta h^J) + E^I_a \xi^a$, where $\xi_a$ is the gaussian noise and $E$ is the inverse of the vielbein for $G_{IJ}$\cite{Susanne}. For $\mathcal{O}(G_N^0)$, we get
\begin{equation}
\begin{split}
G^{IJ} \frac{\delta S^0_{UV}}{\delta h^I} \frac{\delta S^1_{UV}}{\delta h^J} +\frac{1}{2} \left( \frac{ \delta S^1_{UV}}{\delta \phi} \right)^2 - \frac{1}{2}\frac{\delta^2 S^1_{UV}}{\delta \phi^2} &= -(\sqrt{h}) (W(\phi)-\Phi(\phi)R+\frac{1}{2}(\partial_i \phi)^2)\\
=&G^{IJ}\frac{\delta S^0_{IR}}{\delta h^I}\frac{\delta S^1_{IR}}{\delta h^J} +\frac{1}{2} \left( \frac{\delta S^1_{IR}}{\delta \phi} \right)^2 - \frac{1}{2} \frac{\delta^2 S^1_{IR}}{\delta \phi^2}.\label{WDW}
\end{split}
\end{equation}

 Let's derive the exact RG equation from the above equation of motion (\ref{WDW}). We use the right hand side and the left hand side of the equality, and use $\delta S^0_{UV} / \delta h + \delta S^0_{IR} / \delta h = 0$, because $S^0$ is classical, purely gravitational part of the action, and the configuration of the metric $h$ should minimize this. (See \cite{VV}) Remember that now the scalar field $\phi$ is quantized and we can't naively consider its classical trajectory. We have the classical gravity background, and we parameterize the physical scale by the overall factor of the metric. By parameterizing $h_{ij}= e^{2a} \delta_{ij}$, "$-a$" can be considered as the parameter for the RG flow.

\begin{equation}
-\frac{d}{da} S^1 = - G^{IJ} \frac{\delta S^0_{UV}}{\delta h^I} \frac{\delta S^1}{\delta h^J}
=-\left( \frac{1}{2} \frac{\delta S^1}{\delta \phi} -\frac{\delta S^1_{UV}}{\delta \phi}  \right) \frac{\delta S^1}{\delta \phi}
 + \left( \frac{1}{2} \frac{\delta^2 S^1}{\delta \phi^2} -\frac{\delta^2 S^1_{UV} }{\delta \phi^2} \right)\label{HERG1}
\end{equation}

Here we have assumed the behavior of the metric $h$ is determined by the leading order of the action $S_{UV}=S^0_{UV} / 16\pi G_N$, together with (\ref{velocity}).

The above expression is a class of the exact RG equation presented in \cite{Morris},\cite{Morris2},\cite{Oliver}. Please remember, in order to identify the above equation as the exact RG, we needed to identify the boundary value of the bulk field as the dynamical field on the boundary CFT as well. At least it is formally true in some cases of AdS/CFT where we can take free boundary condition, there we integrate over all the possible configuration of the boundary value of the bulk field, and it was treated as a dynamical field on the boundary \cite{MukundHRG}, \cite{WittenNeumann}. Moreover, when we put the boundary at the finite radial distance, the boundary condition of the stochastic string \cite{StochasticSon}, \cite{StochasticdeBoer} and graviton \cite{VV} were identified as the dynamical field on the boundary. However, in order to justify it from the field theory, we would need to show that the effective action can be thought of as a classical action, and that the value of the one-point function can be thought of as the value of the field, and that those fields are the fundamental degree of freedom. We will probably need the discussion about the large N field theory as a classical theory \cite{Yaffe}, \cite{Barbon}, and about the formulation of the gauge theory in terms of Wilson loop \cite{largeN}.


Note that the expression (\ref{HERG1}) can be thought of as a Fokker-Planck equation if we identify $-da$ or $-dr$ as time, and identify probability density as $P= \exp(-S^1)$, and if $S_{UV}$ is nearly constant with respect to $r$ (In the regime of CFT). It implies that it would be also possible to formulate this exact RG equation as a Langevin equation \cite{Gaite}, $ -\partial_a \phi= -\delta S_{UV} / \delta \phi + \xi $, or in terms of the boundary CFT, $S_{UV}$ will be replaced by the generating functional $\Gamma[\phi]$.

\section{From the viewpoint of the Exact RG}

Let's see that the expression we have obtained in the last section is actually the exact RG equation in the field theory. First of all, any RG flow equation are obtained from two steps, coarse-graining the microscopic degree of freedom and rescaling. In the continuum field theory, it is known that the coarse-graining step can be simply performed by redefinition of the field\cite{Oliver}. Therefore, we consider the conformal transformation (scale transformation) of the fields in the CFT, and in addition, we perform extra transformation on scalar field as $\mathcal{F}(\phi) dr$, which will correspond to the coarse-graining. The transformation of the fields will be then,
\begin{equation}
\begin{split}
\phi'&= \phi+ \left( \frac{\delta S^1_{UV}}{\delta \phi} + \mathcal{F}(\phi) \right) dr\\
h'   &= h+ \frac{\delta S^0_{UV}}{\delta h}dr
\label{fieldtransf}
\end{split}
\end{equation}

We now assume that the change of the effective action $S^1$ is solely due to the change of the field, and the functional form of $S^1$ doesn't change, as in the regular derivation of the exact RG. We can write down the Wilsonian effective action $S^1(\phi', h')$ with respect to the new fields $\phi'$, $h'$, and by definition it should satisfy
\begin{equation}
\int \mathcal{D}\phi' \exp(-S^1(\phi', h'))=\int \mathcal{D}\phi \exp(-S^1(\phi, h)).
\end{equation}

Expanding the left hand side with respect to $\phi$, $h$, and collecting terms of the order $\mathcal{O}(dr)$, we get
\begin{equation}
- G^{IJ}\frac{\delta S^0_{UV}}{\delta h^I}\frac{\delta S^1}{\delta h^J}-\frac{\delta S^1_{UV}}{\delta \phi}\frac{\delta S^1}{\delta \phi}
-\mathcal{F}\frac{\delta S^1}{\delta \phi} + \frac{\delta^2 S^1_{UV}}{\delta \phi^2}+ \frac{\delta \mathcal{F}}{\delta \phi}=0.
\end{equation}

If we choose $\mathcal{F}= -\frac{1}{2} \frac{ \delta S^1 }{ \delta \phi} $, this expression is precisely what we have obtained from the Wheeler-DeWitt equation in the bulk,(\ref{HERG1}). The form of the field transformation for the coarse graining $\mathcal{F}$ is intuitively reasonable, because, if we see it in the form of the Langevin equation in the previous section, this term corresponds to the noise, therefore it will look like it is washing out microscopic degree of freedom, as any RG should. It was interpreted in \cite{Gaite} that this noise term corresponds to the contribution from irrelevant operators.

\section{Example: Flow of the double trace operator}

We then still consider the alernative quantization, in which the bulk operators are interpreted as the boundary operator at some RG scale. We mostly follow the discussion in \cite{Polchinski} and \cite{MukundHRG}. We assume that the bulk action is of gaussian form, potential containing only mass term. We take gaussian ansatz for $S_{UV}$ and $S_{IR}$ as a function of the boundary value of the fields $\phi$ and the coordinate $z$. (Note that $z$ can be also understood as the scale parameter from the coordinate dependence of the metric, $h_{ij} \sim z^{-2} \delta_{ij}$.)
The expression of the action for the UV part and the IR part respectively, are
\begin{equation}
\begin{split}
S_{UV}&=C(x,z)+ \frac{1}{2} \int \sqrt{h} \left( F_{UV}(x,z) (\phi-\bar{\phi}(x,z))^2 \right) d^4x \\
S_{IR}&= \frac{1}{2} \int \sqrt{h} \left( F_{IR}(x,z) \phi^2 \right) d^4x .
\end{split}
\end{equation}

We then wish to consider the flow of the double trace operator, which will correspond to the coefficient of the $\phi^2$ term in $S=S_{UV}+S_{IR}$, since, according to the recipe of \cite{dVV}, the sum of the UV and IR action should be identified as the CFT effective action. We will need to see the bulk equation of motion for the evolution of $F=F_{UV}+F_{IR}$, and we can compute $F_{UV}$ and $F_{IR}$ respectively from the Schr\"odinger equation as computed in \cite{Polchinski}. Actually, since we restrict the form of the action to the Gaussian, there is no quantum correction to the double trace operator, (because $(\delta S^2 / \delta \phi^2)$ term will have only constant $\mathcal{O}(\phi^0)$ term) and therefore we can look at Hamilton-Jacobi equation
($\mathcal{H}= z\partial_z S_{UV}=(1/2 \sqrt{h})p^2+(\sqrt{h}/2)( z^2 \delta^{ij} k_i k_j \phi^2 + m^2 \phi^2)$).
 Then we will get\footnote{Note that the below result differ from \cite{Polchinski} since their definition of $F$ includes $\sqrt{h}$ as well.}
\begin{equation}
z \partial_z (\frac{1}{2} z^{-d} F_{UV})=\frac{1}{2}(z^{2-d}k^2 + z^{-d}m^2) - \frac{1}{2} z^{-d}F_{UV}^2.
\end{equation}

This equation can be solved by $F_{UV}= z \partial_z \mathcal{O} / \mathcal{O}$, where $\mathcal{O}$ is the Bessel function which solves
\begin{equation}
z \partial_z (z^{-d} z \partial_z \mathcal{O}) = z^{-d} (z^2 k^2 + m^2) \mathcal{O}.
\end{equation}

The above $F_{UV}$, for $kz  \ll  1$, gives
\begin{equation}
F_{UV} \sim \frac{\Delta_- A_{UV}(k)+ \Delta_+ z^{2\nu}}{A_{UV}(k)+ z^{2\nu}}.
\end{equation}
Note that the behavior of $F_{UV}$ for small/large $z$ is as expected, because it should follow the conformal scaling with $\Delta_-$, $\Delta_+$.
On the other hand, we can also obtain $F_{IR}$ from the condition that the solution will not blow up at $z \rightarrow \infty$, and get
\begin{equation}
F_{IR}=- \frac{z \partial_z (z^{d/2}K_{\nu}(kz))}{z^{d/2}K_{\nu}(kz)}.
\end{equation}

When $k \sim 0$, it can be approximated as $F_{IR} \sim -\Delta_-$.
Therefore, if the identification of the total bulk action as the CFT effective action (as \cite{dVV}) is correct, $F$, sum of $F_{UV}$ and $F_{IR}$, should give the right coefficient for the double trace operator, and it is
\begin{equation}
F= F_{IR}+ F_{UV} \sim \frac{(\Delta_+-\Delta_-)z^{2\nu}}{A_{UV}(k)+z^{2\nu}}.
\end{equation}

 Behavior of $F$ at small $z$, $F \sim z^{2\nu}$, is as expected from the conformal dimention of the coupling of the double trace operator. It is also confirmed that this $F \sim F_{UV}-\Delta_-$ follows the same equation as the flow of the coupling of the double trace operator \cite{MukundHRG}. Here we wish to derive the same equation from a different perspective, from the exact RG flow equation that we have derived.
By rewriting (\ref{HERG1}) with respect to $S$ and $S_{IR}$, and concentrating on $\phi^2$ term, we will get an equation

\begin{equation}
z\partial_z F = F^2 -(d-2\Delta_-)F,\label{doubletrace}
\end{equation}
which is the same equation as in \cite{MukundHRG} up to different definition of $F$ by factor of minus sign. We can see that the derivation of the RG flow equation of the double trace operator naturally arises in our case (summing over UV part and IR part of the action). It still remains to be done what will be the modification of the RG flow equation (\ref{doubletrace}) for general $k$.

\section{Discussion}
Therefore, we gave the holographic definition of the Wilsonian effective action, and derived the exact RG flow equation from the Wheeler-DeWitt equation in the bulk, when we can take free boundary condition for the bulk scalar field. The fact, that we could derive the exact RG from the quantum evolution of the bulk, could be convinced in light of stochastic quantization, which also realize the holographic setup; Firstly, It was proposed that the exact RG equation is equivalent to Stochastic RG equation, which looks like Langevin equation, where the scale plays the role of "time". (See \cite{Gaite}\footnote{The apperence of Langevin equation in the context of RG is not limited to Field Theory. See\cite{Oono}, there they also show that Langevin equation appears as the result of RG}. This RG is basically equivalent to Stochastic Quantization.) and also, in the context of the Stochastic quantization, 
the noise of the Langevin equation along the "extra dimension" on the boundary is corresponding to the Quantum fluctuation of the bulk. Summing up those ingredients, we can convince myself that there is a good reason to suspect the quantum evolution of the bulk is related to the "Stochastic RG", or exact RG of the boundary theory. Partial evidence that the Wheeler-DeWitt equation is related to the exact RG equation also comes from \cite{Polchinski}, where they use the Schr\"odinger equation of the bulk to derive the exact RG. Moreover, in the context of the gauge theory of the boundary, the connection between the exact RG and Schwinger-Dyson equation was made \cite{Hirano}, and in addition, it was proposed that the Schwinger-Dyson equation of the boundary is related to the Wheeler-DeWitt equation of the string theory \cite{Gilad}. This gauge theory discussion also suggests the relationship between the Wheeler-DeWitt equation and the exact RG.

This result will hopefully give a better understanding on how exact RG is related to the quantized gravity, and in relating the holographic RG and Polchinski RG. Also, we wish that there will be more understanding of Holography in terms of Stochastic Quantization as exact RG, which was only implied here.
If this identification is established, we will be able to perform the exact RG in a much simpler way, because many miracles, such as no need of gauge-fixing, no need of putting the cutoff, simplification for large N and natural apperance of supersymmetry so on, happens in the stochastic quantization. Note that the same analysis would be applicable to Horava gravity, since they are constructed out of the stochastic quantization, and the stochastic quantization is equivalent to a class of exact RG. It was explained that the structure of stochastic quantization enables the simpler proof of its renormalizability\cite{Susanne}, and we expect that might apply to other general holographic models. Further it was implied that the quantized gravity in the bulk will be written as Langevin equation, which might give more insights into microscopic picture for thermodynamics of spacetime\cite{Jacobson}, \cite{Entropic}.

\section{Update on recent research}
Since this note was originally written more than 10 years ago, there have been a number of recent studies \cite{Sin}\cite{Balasubramanian},\cite{Leigh},\cite{Sathiapalan1},\cite{Sathiapalan2},\cite{Sathiapalan3},\cite{Sathiapalan4},\cite{Sathiapalan5}\cite{Sathiapalan6}, that need to be mentioned.
Some of them \cite{Sin}\cite{Balasubramanian} view the holographic setting in a similar way by splitting to UV part and IR part, although they are based on standard quantization and consider the gravity action after it is integrated over quantum fluctuation.
A series of studies \cite{Sathiapalan1},\cite{Sathiapalan2},\cite{Sathiapalan3},\cite{Sathiapalan4},\cite{Sathiapalan5},\cite{Sathiapalan6} start from the exact RG equation and express transition amplitudes between wave functions of different scale as $\int \mathcal{D}x(t) e^{-S_{bulk}}$, where $x(t)$ denotes fields at different scale $t$, and regard $S_{bulk}$ as the bulk action. They also mention that some of their derivation corresponds to alternative quantization\cite{Sathiapalan2}, and the relationship between Holographic exact RG and Stochastic Quantization\cite{Sathiapalan5}.
The goal of relating the exact RG to the bulk quantum gravity is the same, although they are working on deriving the bulk theory without using explicitly holographic setting.
This note is different in working with alternative quantization and using holographic setting, deriving the exact RG equation on the boundary from the evolution of the wave function in the bulk, dealing with the counterterm.
There are also a number of studies that relate Stochastic Quantization and Holography\cite{SQ1},\cite{SQ2},\cite{SQ3},\cite{SQ4},\cite{SQ5},\cite{SQ6}. In particular, \cite{SQ2},\cite{SQ3},\cite{SQ4},\cite{SQ5},\cite{SQ6} identify radial evolution in the bulk to Stochastic Quantization, which is the exact RG on the boundary, and also derives the evolution of double trace operator, so they are practically very similar to this note. However, their derivation is different from the one in this note, in particular how to subtract the counterterm.
To note, in addition to the research above, we were recently informed that there is another study that mentions the resemblance between the exact RG and Fokker-Planck equation\cite{Hatta}.

\section*{Acknowledgements}
I would like to thank Gilad Lifschitz, Akio Sugamoto and So Katagiri for helpful comments, and those who answered my questions on holography and exact RG.


\begin{thebibliography}{99}

\bibitem{dVV}
  J.~de Boer, E.~P.~Verlinde and H.~L.~Verlinde,
  ``On the holographic renormalization group,''
  JHEP {\bf 0008} (2000) 003
  [arXiv:hep-th/9912012].

\bibitem{VV}
  E.~P.~Verlinde and H.~L.~Verlinde,
  ``RG-flow, gravity and the cosmological constant,''
  JHEP {\bf 0005}, 034 (2000)
  [arXiv:hep-th/9912018].

\bibitem{Polchinski}
  I.~Heemskerk and J.~Polchinski,
  ``Holographic and Wilsonian Renormalization Groups,''
  arXiv:1010.1264 [hep-th].

\bibitem{Parisi}
  G.~Parisi and Y.~s.~Wu,
  ``Perturbation Theory Without Gauge Fixing,''
  Sci.\ Sin.\  {\bf 24}, 483 (1981).

\bibitem{Damgaard}
  P.H.Damgaard and H.Huffel,
  ``STOCHASTIC QUANTIZATION,''
{\it  SINGAPORE, SINGAPORE: WORLD SCIENTIFIC (1988) 496p},

\bibitem{Sakai}
  M.~Fukuma, S.~Matsuura and T.~Sakai,
  ``Holographic renormalization group,''
  Prog.\ Theor.\ Phys.\  {\bf 109}, 489 (2003)
  [arXiv:hep-th/0212314].

\bibitem{Morris}
  T.~R.~Morris,
  ``The Exact renormalization group and approximate solutions,''
  Int.\ J.\ Mod.\ Phys.\  A {\bf 9}, 2411 (1994)
  [arXiv:hep-ph/9308265].

\bibitem{Morris2}
    T.~R.~Morris,
  ``A manifestly gauge invariant exact renormalization group,''
  arXiv:hep-th/9810104.

\bibitem{Morris3}
    J.~I.~Latorre and T.~R.~Morris,
  ``Exact scheme independence,''
  JHEP {\bf 0011}, 004 (2000)
  [arXiv:hep-th/0008123].

\bibitem{Oliver}
  O.~J.~Rosten,
  ``Fundamentals of the Exact Renormalization Group,''
  arXiv:1003.1366 [hep-th].

\bibitem{Witten2}
  I.~R.~Klebanov and E.~Witten,
  ``AdS/CFT correspondence and symmetry breaking,''
  Nucl.\ Phys.\  B {\bf 556}, 89 (1999)
  [arXiv:hep-th/9905104].

\bibitem{Muck}
  W.~Mueck and K.~S.~Viswanathan,
  ``Regular and irregular boundary conditions in the AdS/CFT  correspondence,''
  Phys.\ Rev.\  D {\bf 60}, 081901 (1999)
  [arXiv:hep-th/9906155].

\bibitem{Witten}
  E.~Witten,
  ``Multi-trace operators, boundary conditions, and AdS/CFT correspondence,''
  arXiv:hep-th/0112258.

\bibitem{WittenNeumann}
  E.~Witten,
  ``SL(2,Z) action on three-dimensional conformal field theories with Abelian symmetry,''
  arXiv:hep-th/0307041.

\bibitem{MukundHRG}
  T.~Faulkner, H.~Liu, M.~Rangamani,
  ``Integrating out geometry: Holographic Wilsonian RG and the membrane paradigm,''
  [arXiv:1010.4036 [hep-th]].

\bibitem{MukundOfer}
  O.~Aharony, D.~Marolf and M.~Rangamani,
  ``Conformal field theories in anti-de Sitter space,''
  JHEP {\bf 1102}, 041 (2011)
  [arXiv:1011.6144 [hep-th]].

\bibitem{Ross}
  D.~Marolf and S.~F.~Ross,
  ``Boundary conditions and new dualities: Vector fields in AdS/CFT,''
  JHEP {\bf 0611}, 085 (2006)
  [arXiv:hep-th/0606113].

\bibitem{StochasticdeBoer}
  J.~de Boer, V.~E.~Hubeny, M.~Rangamani {\it et al.},
  ``Brownian motion in AdS/CFT,''
  JHEP {\bf 0907}, 094 (2009).
  [arXiv:0812.5112 [hep-th]].
  
\bibitem{StochasticSon}
  D.~T.~Son, D.~Teaney,
  ``Thermal Noise and Stochastic Strings in AdS/CFT,''
  JHEP {\bf 0907}, 021 (2009).
  [arXiv:0901.2338 [hep-th]].

\bibitem{PolchinskiRG}
  J.~Polchinski,
  ``Renormalization And Effective Lagrangians,''
  Nucl.\ Phys.\  B {\bf 231}, 269 (1984).


\bibitem{GiladSchwinger}
  G.~Lifschytz, S.~D.~Mathur and M.~Ortiz,
  ``A Note on the semiclassical approximation in quantum gravity,''
  Phys.\ Rev.\  D {\bf 53}, 766 (1996)
  [arXiv:gr-qc/9412040].

\bibitem{Nishioka}
  T.~Nishioka,
  ``Horava-Lifshitz Holography,''
  Class.\ Quant.\ Grav.\  {\bf 26}, 242001 (2009)
  [arXiv:0905.0473 [hep-th]].

\bibitem{Susanne}
  D.~Orlando and S.~Reffert,
  ``On the Renormalizability of Horava-Lifshitz-type Gravities,''
  Class.\ Quant.\ Grav.\  {\bf 26}, 155021 (2009)
  [arXiv:0905.0301 [hep-th]].

\bibitem{Yaffe}
  L.~G.~Yaffe,
  ``Large N Limits As Classical Mechanics,''
  Rev.\ Mod.\ Phys.\  {\bf 54}, 407 (1982).

\bibitem{Barbon}
  J.~L.~F.~Barbon,
  ``Multitrace AdS/CFT and master field dynamics,''
  Phys.\ Lett.\  B {\bf 543}, 283 (2002)
  [arXiv:hep-th/0206207].

\bibitem{largeN}
  Y.~Makeenko,
  ``Large-N gauge theories,''
  arXiv:hep-th/0001047.

\bibitem{Gaite}
  J.~C.~Gaite,
  ``Stochastic formulation of the renormalization group: Supersymmetric
  structure and topology of the space of couplings,''
  J.\ Phys.\ A  {\bf 37}, 10409 (2004)
  [arXiv:hep-th/0404212].

\bibitem{Oono}
Some implications of renormalization group theoretical ideas to statistics
Physica D: Nonlinear Phenomena, Volume 205, Issue 1-4, June 2005, Pages 207-214
Rajaram, S.; Taguchi, Y.h.; Oono, Y.

\bibitem{Hirano}
  S.~Hirano,
  ``Exact renormalization group and loop equation,''
  Phys.\ Rev.\  D {\bf 61}, 125011 (2000)
  [arXiv:hep-th/9910256].

\bibitem{Gilad}
  G.~Lifschytz and V.~Periwal,
  ``Schwinger-Dyson = Wheeler-DeWitt: Gauge theory observables as bulk
  operators,''
  JHEP {\bf 0004}, 026 (2000)
  [arXiv:hep-th/0003179].

\bibitem{Jacobson}
  T.~Jacobson,
  ``Thermodynamics of space-time: The Einstein equation of state,''
  Phys.\ Rev.\ Lett.\  {\bf 75}, 1260 (1995)
  [arXiv:gr-qc/9504004].

\bibitem{Entropic}
  E.~P.~Verlinde,
  ``On the Origin of Gravity and the Laws of Newton,''
  arXiv:1001.0785 [hep-th].


\bibitem{Sin}
S.~J.~Sin and Y.~Zhou,
``Holographic Wilsonian RG Flow and Sliding Membrane Paradigm,''
JHEP \textbf{05}, 030 (2011)
doi:10.1007/JHEP05(2011)030
[arXiv:1102.4477 [hep-th]].

\bibitem{Balasubramanian}
V.~Balasubramanian, M.~Guica and A.~Lawrence,
``Holographic Interpretations of the Renormalization Group,''
JHEP \textbf{01}, 115 (2013)
doi:10.1007/JHEP01(2013)115
[arXiv:1211.1729 [hep-th]].

\bibitem{Leigh}
R.~G.~Leigh, O.~Parrikar and A.~B.~Weiss,
``Exact renormalization group and higher-spin holography,''
Phys. Rev. D \textbf{91}, no.2, 026002 (2015)
doi:10.1103/PhysRevD.91.026002
[arXiv:1407.4574 [hep-th]].

\bibitem{Sathiapalan1}
B.~Sathiapalan and H.~Sonoda,
``A Holographic form for Wilson's RG,''
Nucl. Phys. B \textbf{924}, 603-642 (2017)
doi:10.1016/j.nuclphysb.2017.09.018
[arXiv:1706.03371 [hep-th]].

\bibitem{Sathiapalan2}
B.~Sathiapalan and H.~Sonoda,
``Holographic Wilson's RG,''
Nucl. Phys. B \textbf{948}, 114767 (2019)
doi:10.1016/j.nuclphysb.2019.114767
[arXiv:1902.02486 [hep-th]].

\bibitem{Sathiapalan3}
B.~Sathiapalan,
``Holographic RG and Exact RG in O(N) Model,''
Nucl. Phys. B \textbf{959}, 115142 (2020)
doi:10.1016/j.nuclphysb.2020.115142
[arXiv:2005.10412 [hep-th]].

\bibitem{Sathiapalan4}
P.~Dharanipragada, S.~Dutta and B.~Sathiapalan,
``Bulk gauge fields and holographic RG from exact RG,''
JHEP \textbf{23}, 174 (2020)
doi:10.1007/JHEP02(2023)174
[arXiv:2201.06240 [hep-th]].

\bibitem{Sathiapalan5}
P.~Dharanipragada, S.~Dutta and B.~Sathiapalan,
``Aspects of the map from exact RG to holographic RG in AdS and dS,''
Mod. Phys. Lett. A \textbf{37}, no.37n38, 2250235 (2022)
doi:10.1142/S0217732322502352
[arXiv:2301.13605 [hep-th]].

\bibitem{Sathiapalan6}
P.~Dharanipragada and B.~Sathiapalan,
``Holographic RG from ERG: Locality and General Coordinate Invariance in the Bulk,''
[arXiv:2306.07442 [hep-th]].

\bibitem{SQ1}
D.~S.~Mansi, A.~Mauri and A.~C.~Petkou,
``Stochastic Quantization and AdS/CFT,''
Phys. Lett. B \textbf{685}, 215-221 (2010)
doi:10.1016/j.physletb.2010.01.033
[arXiv:0912.2105 [hep-th]].

\bibitem{SQ2}
J.~H.~Oh and D.~P.~Jatkar,
``Stochastic quantization and holographic Wilsonian renormalization group,''
JHEP \textbf{11}, 144 (2012)
doi:10.1007/JHEP11(2012)144
[arXiv:1209.2242 [hep-th]].

\bibitem{SQ3}
D.~P.~Jatkar and J.~H.~Oh,
``Stochastic quantization of conformally coupled scalar in AdS,''
JHEP \textbf{10}, 170 (2013)
doi:10.1007/JHEP10(2013)170
[arXiv:1305.2008 [hep-th]].

\bibitem{SQ4}
J.~H.~Oh,
``First-order formalism of holographic Wilsonian renormalization group: Langevin equation,''
J. Korean Phys. Soc. \textbf{79}, no.10, 903-917 (2021)
doi:10.1007/s40042-021-00320-x
[arXiv:2110.05013 [hep-th]].

\bibitem{SQ5}
G.~Kim, J.~s.~Chae, W.~Shin and J.~H.~Oh,
``Stochastic quantization and holographic Wilsonian renormalization group of scalar theory with generic mass, self-interaction and multiple trace deformation,''
Int. J. Mod. Phys. A \textbf{38}, no.21, 2350114 (2023)
doi:10.1142/S0217751X23501142
[arXiv:2305.18920 [hep-th]].

\bibitem{SQ6}
J.~H.~Lee and J.~H.~Oh,
``Stochastic quantization and holographic Wilsonian renormalization group of conformally coupled scalar in AdS$_{4}$,''
J. Korean Phys. Soc. \textbf{83}, no.9, 665-674 (2023)
doi:10.1007/s40042-023-00926-3
[arXiv:2308.10010 [hep-th]].

\bibitem{Hatta}
Y.~Hatta and T.~Kunihiro,
``Renormalization group method applied to kinetic equations: Roles of initial values and time,''
Annals Phys. \textbf{298}, 24-57 (2002)
doi:10.1006/aphy.2002.6234
[arXiv:hep-th/0108159 [hep-th]].


\end{thebibliography}
\end{document}